# Crystallite size and microstrain in the structure of $SrTiO_3$ formed by magnetron deposition with and without $O_2$ flow through the deposition chambre

**Zdeněk Jansa[1)], Štěpánka Jansová[1)], Lucie Nedvědová[1)], Ján Minár[1)]**

[1)] *New Technologies Research Centre, University of West Bohemia in Pilsen, Pilsen*

## Introduction

The minerals, commonly called Perovskites, were discovered in the Ural Mountains in 1839. The perovskite discovered in 1839 had the chemical formula $CaTiO_3$. It gave its name to a newly formed group of compounds, the Perovskites. Today, this group contains many hundreds of compounds. These new compounds have a chemical formula derived from or very similar to the original formula, ABX3. The most abundant mineral in solid form on earth belongs to the Perovskite group - it is Bridgmanite with the formula (Fe,Mg)$SiO_3$.[1]

Perovskite compounds are ionic in nature. In the $ABX_3$ formula above, A and B represent two cations, where A is a large cation and B is a medium cation. X is a small anion. The overall ionic structure must be neutral and therefore if we describe the charges on the individual ions qa, qb and qx, then the equation for the neutral configuration will be:

$$q_a + q_b = -3q_x$$

Transition metal oxides, which have a perovskite structure, have received much attention in recent decades. This is because of the very suitable properties that can be used in various industries. These are large negative megnetoresistance, multiferroicity (the material is ferroelectric, thermoelectric, pyroelectric, dielectric), excellent optical properties and others for which these materials are suitable as capacitors, random access memories, tunable microwave devices, capacitors, displays, piezoelectric devices, sensors, actuators, transducers and wireless communication[2,3].

One of the fields where the properties of Perovskites can and have already been applied is in the power industry. Experimental studies of the last few years have reliably demonstrated that, using simple modifications, perovskite oxides can be used in applications utilizing direct sunlight and photocatalytic applications. The original strontium titanate oxide $SrTiO_3$ (abbreviated STO) is only able to use the UV component of incident radiation and is inactive in visible light, remaining transparent to visible light. The reason for this state is its wide band gap, which at room temperature has a value of 3.2 - 3.25 eV. Studies have shown that doping the structure of these oxides with transition metals (TM) can cause a shift in the valence and/or conduction band. This is because of the 3d dopant bands that create new energy levels in the mentioned band gap, effectively reducing it[3,4].

One of the elements of the TM group is nickel. According to some studies, it is likely that the Ni ion occurs in the cubic structure of STO in the form of Ni2+ and substitutes the Ti4+ sites. In the case of using this dopant, the shift of the absorption edge of STO:Nix relative to NiO was found to be 1.1 eV.

The present work investigates the differences in the structure of STO:Nix prepared by magnetron deposition method with different dopant amount settings. At the same time, the difference in structure was observed when prepared under vacuum with Ar working gas and when O2 was flowed through the deposition chamber. The premise of this experiment was to verify the formation of oxygen vacancies in the STO structure on which the photocatalytic phenomenon could occur[6,7].

## Samples and methods

All samples were prepared in a magnetron deposition chamber. The TF 600 BOC Edward deposition apparatus was used for the growth of all STO thin films investigated. The deposition system allows for the preparation of thin films by magnetron sputtering using two magnetrons, the first connected to a high frequency source (HF - 13.56 MHz) and the second to a DC source. Two series of samples were prepared in the chamber. The deposition chamber in the case of the first series was depleted to a base pressure of 2x10-4 Pa and filled with argon working gas. In the second case, the oxygen flow rate through the deposition chamber was set to 1.75 sccm.

The procedure of the experimental methods was identical for both series. After fabrication, all samples were first subjected to X-ray phase analysis. The samples were measured by X-ray diffraction (XRD) on a Panalytical X'Pert automatic powder diffractometer. The samples were measured using both possible geometries available in the goniometer of the machine. The first is the symmetric θ-θ geometry. The second is an asymmetric ω-2θ geometry with a fixed X-ray tube angle of ω = 0.5°. This method is referred to by the English term Grazing incidence Diffraction (GID). This angle allows very little penetration of the primary beam into the sample under investigation with a large irradiated sample area. Both symmetries were used in the range of 20° to 75° 2θ. In the initial stage, the samples were remeasured by GID in their original as-deposited state, then the samples were subjected to an in-situ experiment in a high-temperature chamber with a maximum temperature of 900°C. Due to the time-consuming nature of the experiment, the samples were measured in the high-temperature chamber using the θ-θ method. Then, the samples were measured by GID after removal from the high temperature chamber and then the data were analyzed.

The second experimental method used was scanning electron microscopy. The chemical composition of the samples and observation of the surface morphology of the samples and their fractures was performed on a JSM-7600F electron microscope from JEOL.

The third method was X-ray photoemission electron spectroscopy (XPS). The experiments on the selected samples were carried out in the working chamber. In this chamber, an ultra-high vacuum (base pressure ≤ 5×10-8 Pa) and a Phoibos 150 hemispherical analyzer were used throughout the experiments. The photon source was a Mg and Al X-ray tube.

## X-ray phase analysis

Seven samples were generated using magnetron deposition. The samples were prepared by magnetron sputtering from an STO target fitted with different numbers of nickel pellets. Some of the samples were prepared in an atmosphere of 100% argon. The other part of the samples was prepared in an atmosphere consisting of a mixture of 75% argon and 25% oxygen. The samples are labeled $A_0$ *with* $O_2$ - sample without dopant used with $O_2$ flow rate, $A_1$ sample with one Ni pellet, *$A_1$ with $O_2$* - sample with one Ni pellet with $O_2$ flow rate. Analogously, the designation is for samples with two and three Ni pellets.

Measurements of nickel-doped strontium titanate samples in the as-deposited state were performed with the same conclusion for all samples. Only the amorphous phase and the incipient crystallization of some phases can be distinguished in the diffraction record. It was not possible to determine the necessary crystallographic data and it was therefore clear that the evaluation was always important only in the as-deposited state.

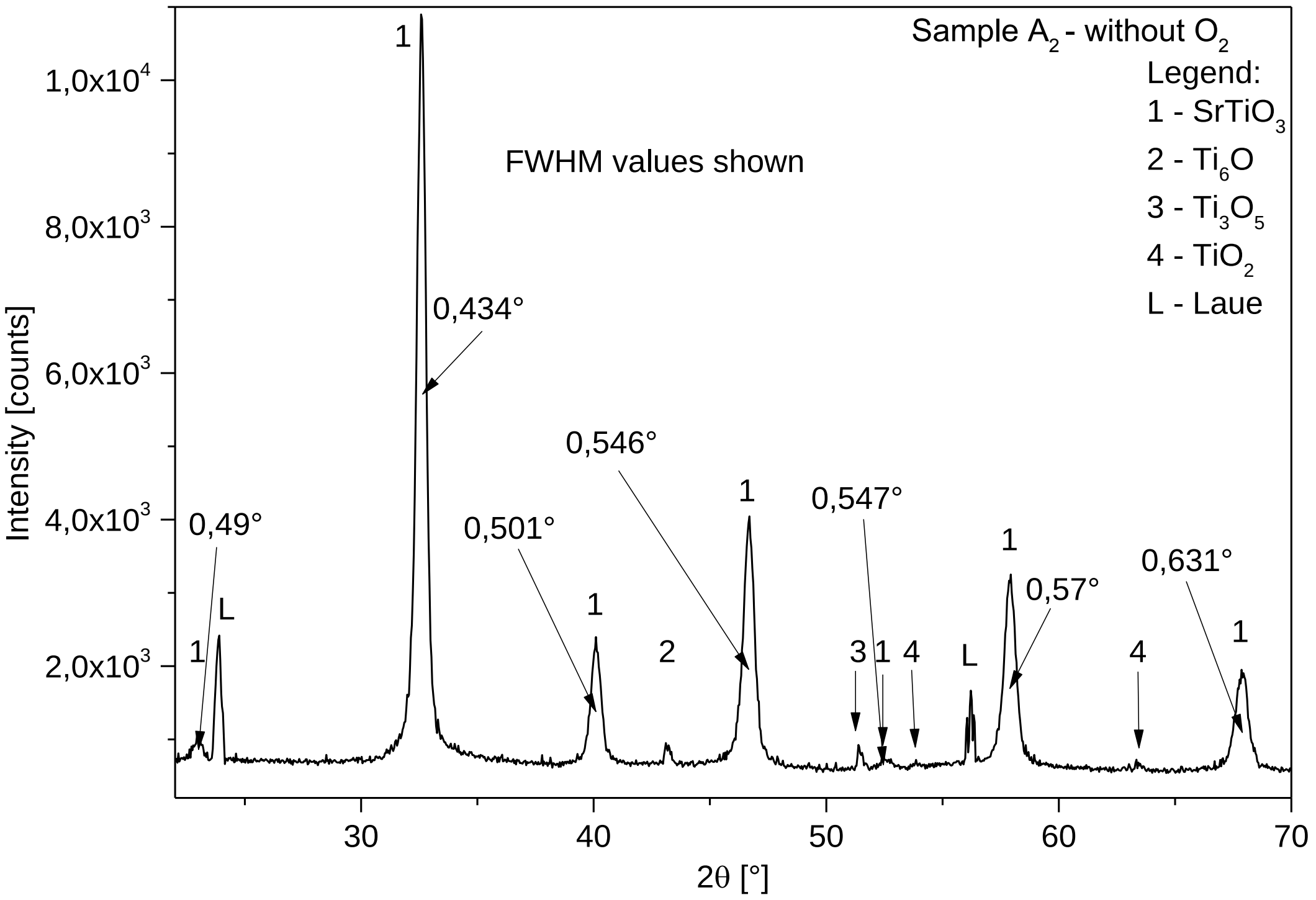

Figure 1 - Diffractogram evaluation of sample *$A_2$*, created without $O_2$ flow.

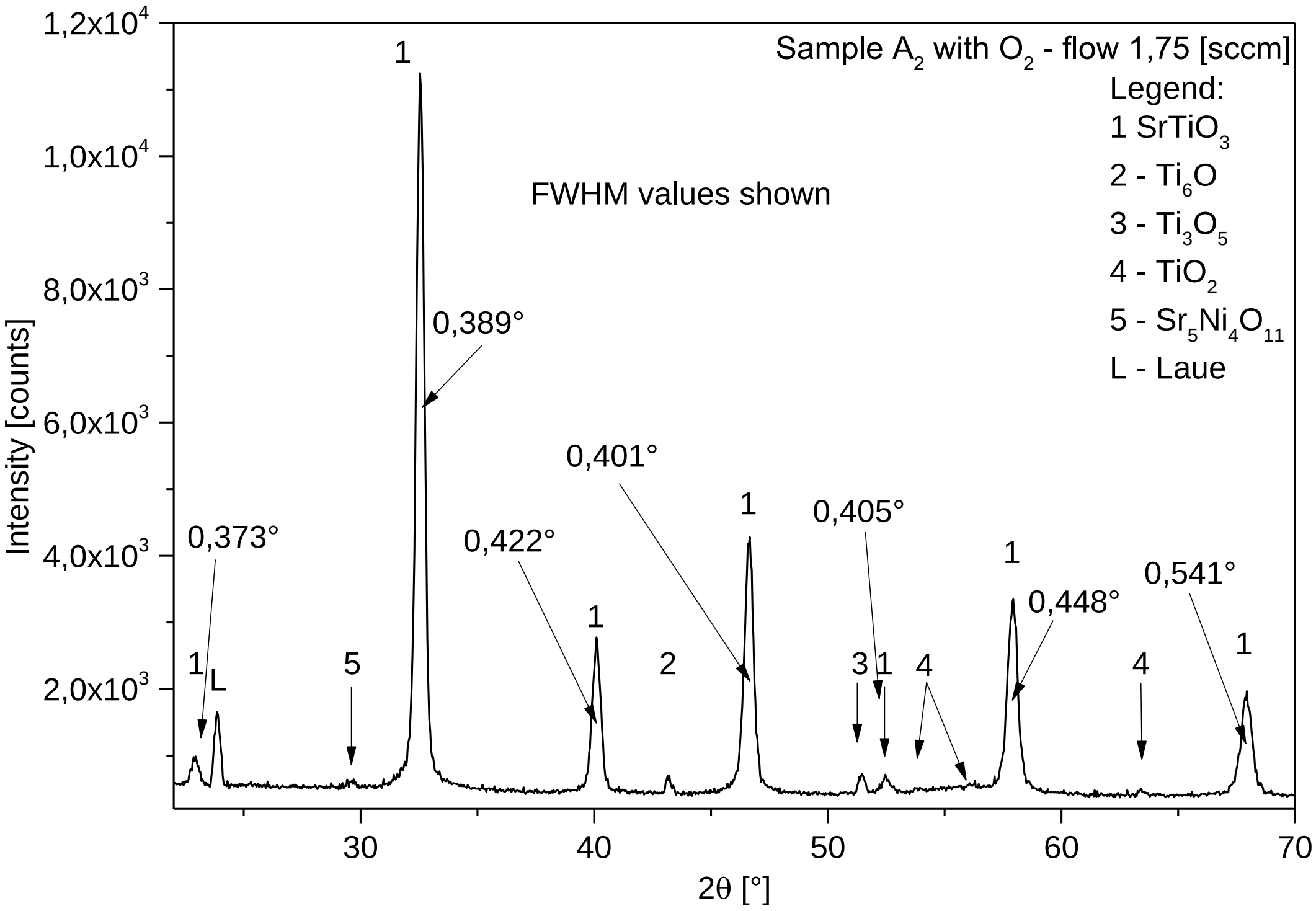

Figure 2 - Diffractogram evaluation of sample $A_2$ *with* $O_2$, $O_2$ flow rate = 1.75 [sccm].

In Figures 1 and 2, the FWHM values are shown for the diffraction maxima of the STO planes. These values illustrate the effect of $O_2$ flow through the deposition chamber on its width.

By analyzing the diffractogram from the measurement of sample *A2*, i.e. the sample without $O_2$ flow doped with 2 nickel pellets, it was found that the following phases occur in the sample structure. The majority phase is the cubic STO phase. The oxide phases of titanium with different stoichiometries are present in the sample structure as accompanying phases. In the investigated diffractogram in Figure 1, the FWHM values of the STO phase are shown for all diffracting crystallographic planes present.

Figure 2 shows the analysis of the measurement record of sample $A_2$ with $O_2$, which is a sample deposited with $O_2$ flow and 2 nickel pellets. As can be seen at first glance, the two diffractograms of samples $A_2$ and $A_2$ with $O_2$ are very similar. The only difference is in the identification of the very weak $Sr_5Ni_4O_{11}$ phase. However, its amount is only slightly above the detection limit of the method used. Also for this sample, FWHM values are given for all diffracting planes of the STO phase. As can be seen, the air flow had a demonstrable effect on the crystallite sizes, where a narrowing of the diffraction lines can be seen.

The effect on the magnitude of the FWHM and on the intensity of the diffraction maximum of a given crystallographic plane can be seen in Table 1, where the data for the (110) and (220) planes of all samples are compared, and in Figure 3, where the state of the (110) plane from the above-quoted table is shown graphically. Two basic mechanisms are evident here. For the samples where $O_2$ flow was used in deposition, there is first an increase from an initial value of 0.3759° ($A_0$ *with* $O_2$ - no pellet) to a value of 0.405° ($A_1$ *with* $O_2$ - 1 Ni pellet) and then a gradual decrease to a value of 0.3694° ($A_3$ *with* $O_2$ - 3 Ni pellets). In contrast, for the sample without $O_2$ flow, there was only an increase in FWHM from the initial 0.3529° ($A_1$ - 1 Ni pellet) to 0.5394° ($A_3$ - 3 Ni pellets). For the intensities, the course of the maximum is opposite to the magnitude of the FWHM.

Figure 38 shows this effect graphically. All samples that were produced by pulsed magnetron deposition with $O_2$ flow through the chamber are shown by solid lines. Samples where the chamber was filled with argon only are indicated by a solid line. The line of the plane (110) from the used standard 01-084-0443, which was identical for all samples, is inserted in the graph.

Table 1 - Comparison of FWHM and maximum intensity of crystallographic planes (110) and (220) relative to $O_2$ flow rate and number of nickel dopant pellets. An undoped STO sample without $O_2$ (Pure STO) is shown in the table for comparison.

| Sample | $O_2$ flow [sccm] | Amount of pellets | FWHM [°] | | Intensity [counts] | |
|---|---|---|---|---|---|---|
| | | | (110) | (220) | (110) | (220) |
| Pure STO | 0 | 0 | 0,4317 | 0,5458 | 8978 | 1571 |
| $A_0$ with $O_2$ | 1,75 | 0 | 0,3759 | 0,416 | 5987 | 712 |
| $A_1$ | 0 | 1 | 0,3529 | 0,3977 | 12959 | 2400 |
| $A_1$ with $O_2$ | 1,75 | 1 | 0,405 | 0,57 | 8802 | 1285 |
| $A_2$ | 0 | 2 | 0,4338 | 0,6306 | 9274 | 1349 |
| $A_2$ with $O_2$ | 1,75 | 2 | 0,4013 | 0,541 | 10189 | 1491 |
| $A_3$ | 0 | 3 | 0,5394 | 0,8595 | 5266 | 819 |
| $A_3$ with $O_2$ | 1,75 | 3 | 0,3694 | 0,4934 | 11173 | 1725 |

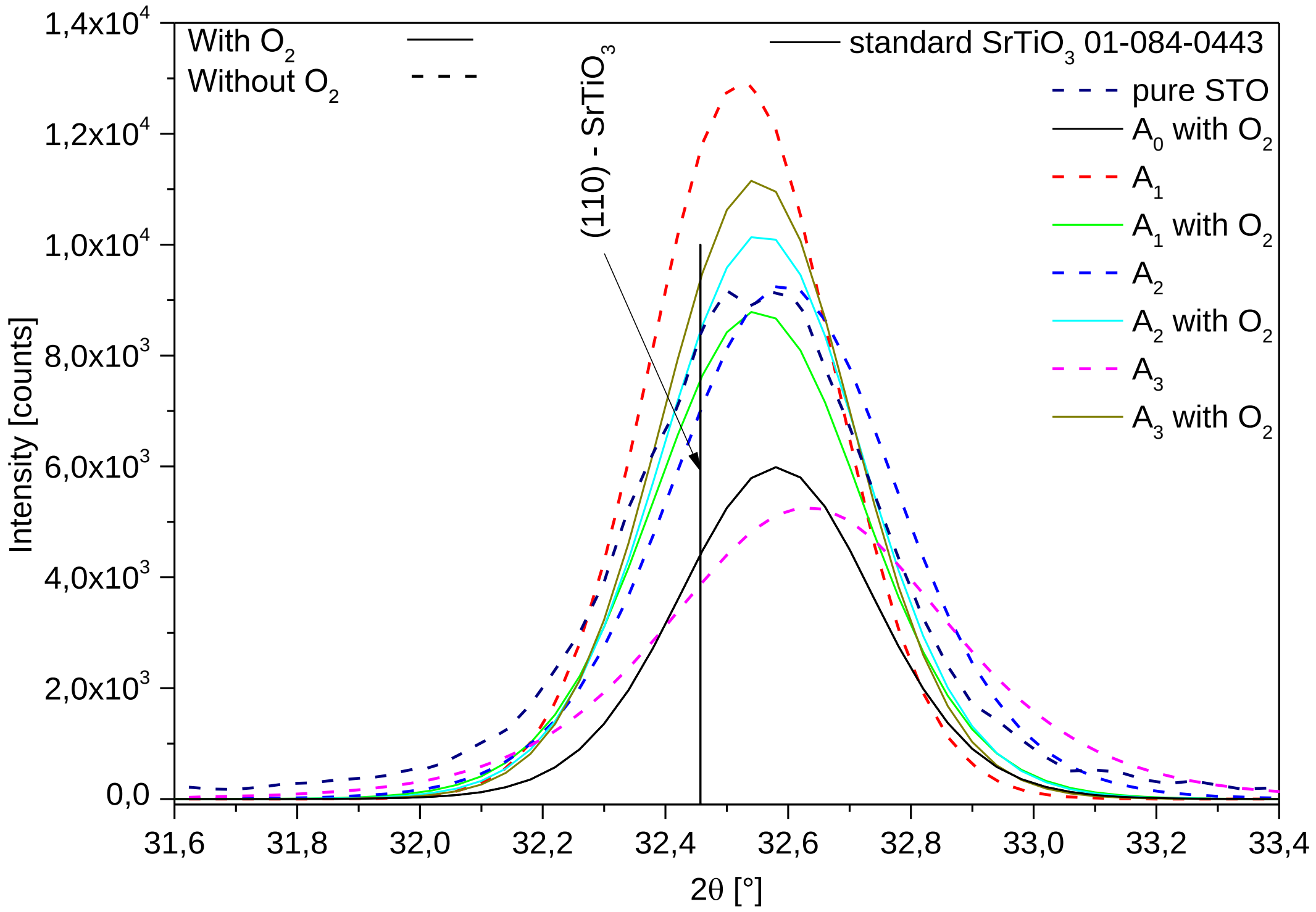

Figure 3 - Dependence of the amount of nickel dopant on the stress generation in the nickel-doped STO lattice and the shift of the diffraction line of the (110) plane - samples $A_1$ to $A_3$ *with* $O_2$.

Due to the demonstrated effect of $O_2$ flow rate on the changes in diffraction line parameters, structural data were calculated for two samples of the second series (*$A_2$* and *$A_2$ with $O_2$*). The results for both samples, namely crystallite size and microdeformation, are graphically presented in Figures 4(a) and (b). Here, a marked difference in the crystallite sizes can be seen.

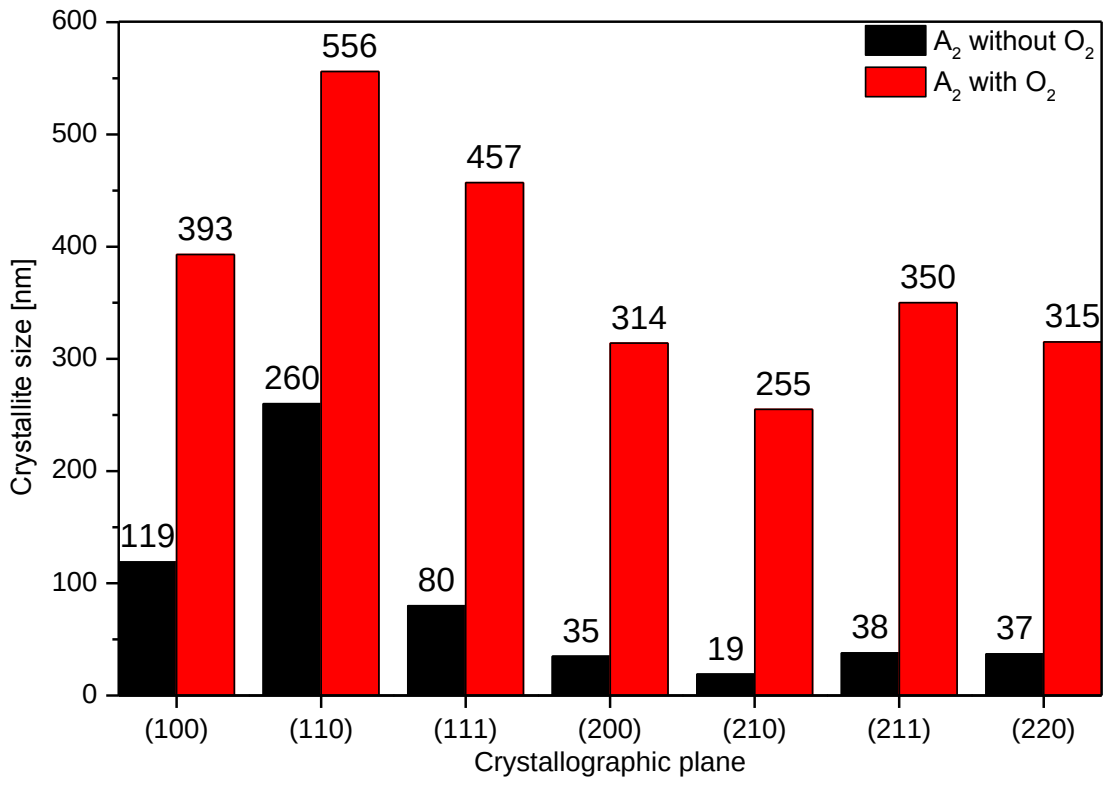

a)

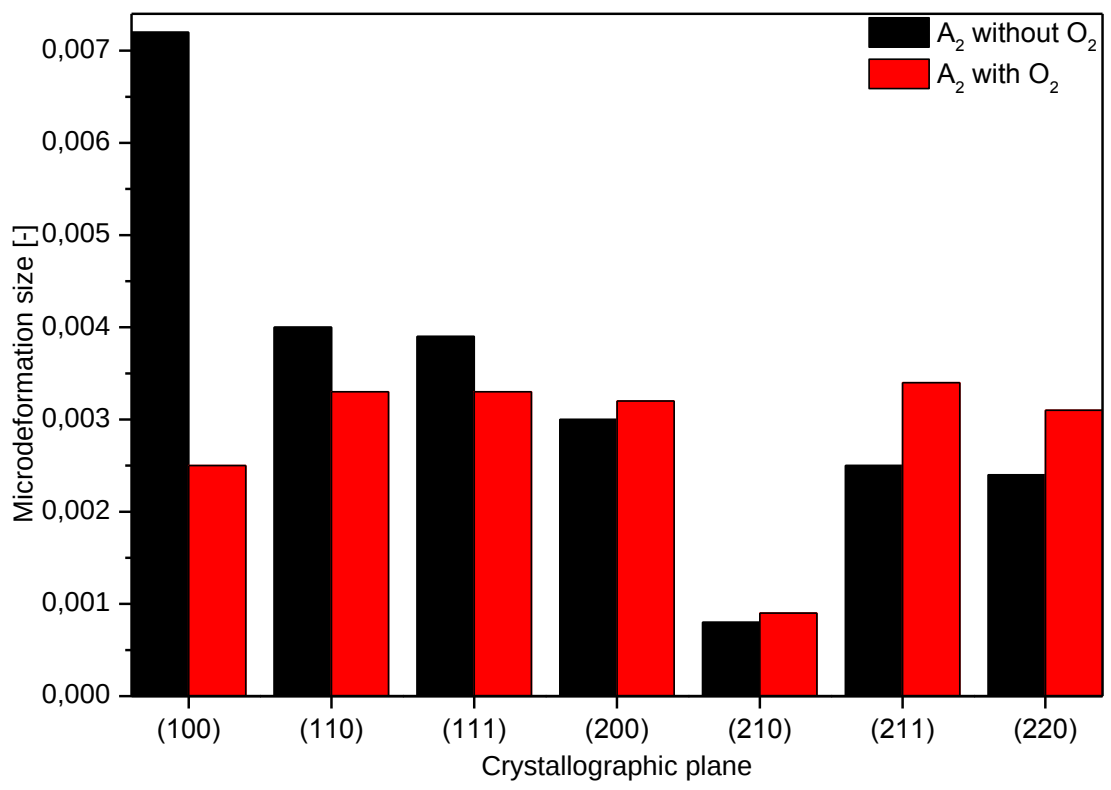

b)

Figure 4 - Graphical representation of the dependence of the amount of dopant and the presence of $O_2$ flow of samples *$A_2$* and *$A_2$ with $O_2$* on a) the size of the crystallites and b) the size of the microdeformations inside the lattice.

**Scanning electron microscopy (SEM)**

Surface examination and topographical data acquisition on nickel-doped STO thin films was performed on all samples. The samples were cleaned in an ultrasonic cleaner before being placed in the vacuum chamber of the SEM, then immersed in liquid nitrogen for 10 seconds and then fractured. The surface quality of the thin layer, possible surface defects and the condition of the fracture edge of the thin layer were examined for the samples.

Sample *$A_2$ with $O_2$*

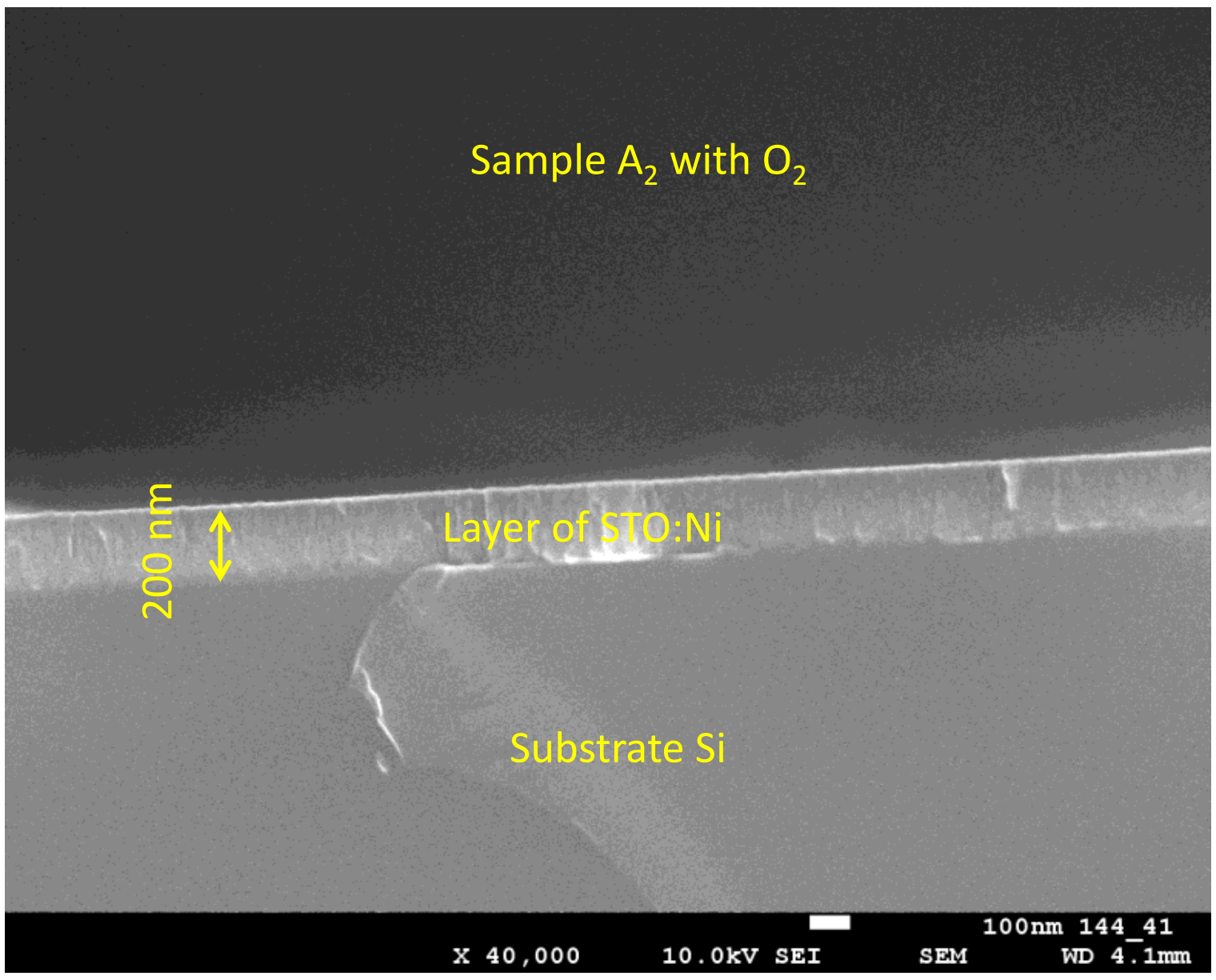

Figure 5 - Perpendicular view of the fracture surface of the *$A_2$ with $O_2$*, magnification 40 000x, layer thickness 200 nm.

Figures 5 and 6 show images of the $A_2$ sample with $O_2$. The first figure is a perpendicular view of the fracture surface of the sample. From this view, the layer thickness (200nm) can be determined and the column texture of the layer growth can be seen. The epitaxy of this sample to the Si substrate is good. In the second image, it can be seen that the sample has a cracked surface, and the peaks of the various columnar growth formations of the sample can also be seen. The reason for the surface cracking of the samples will be further studied to find a way to remove these cracks.

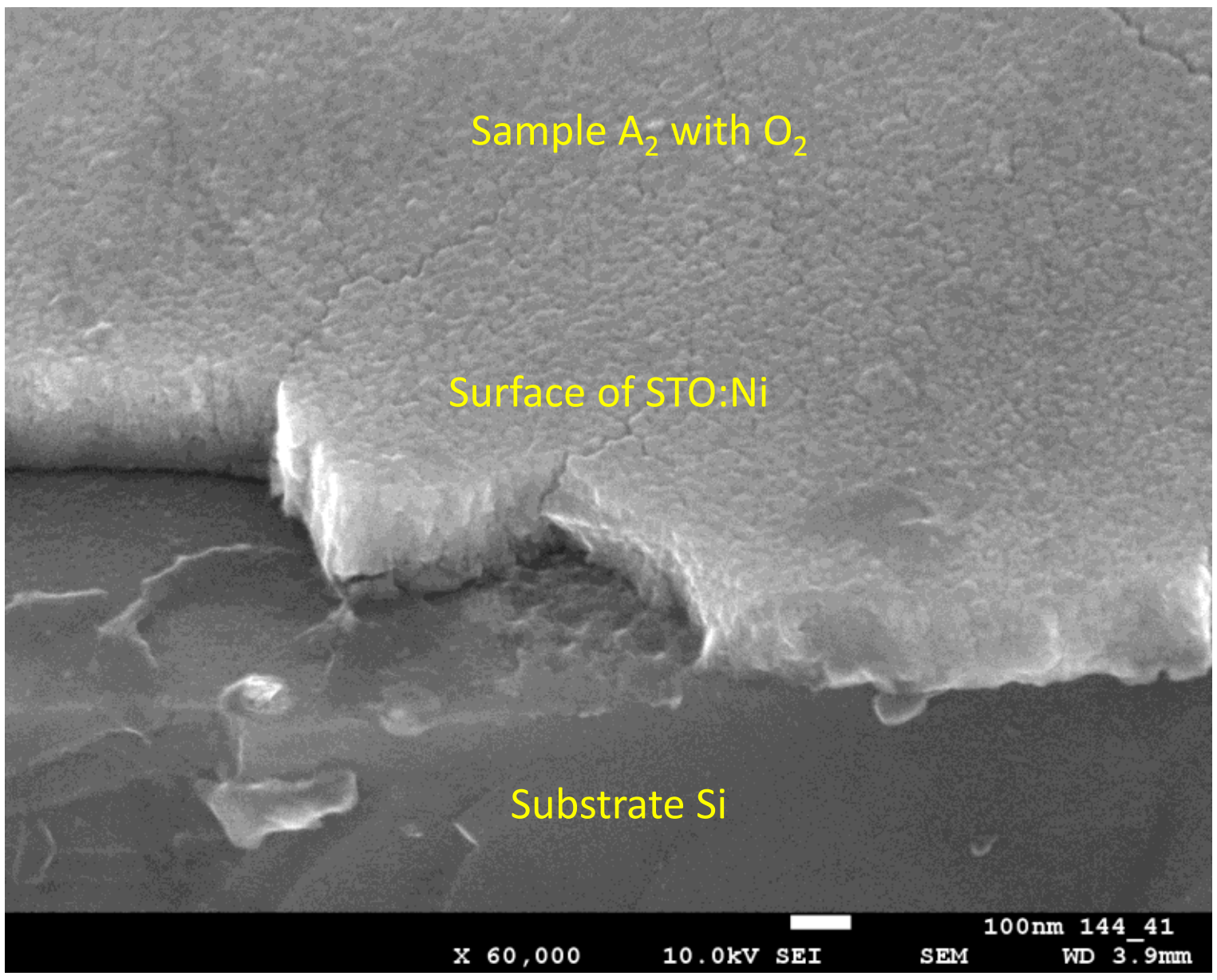

Figure 6 - 45° view, magnification 60 000x, visible cracks in the surface of the thin layer.

$A_3$ sample

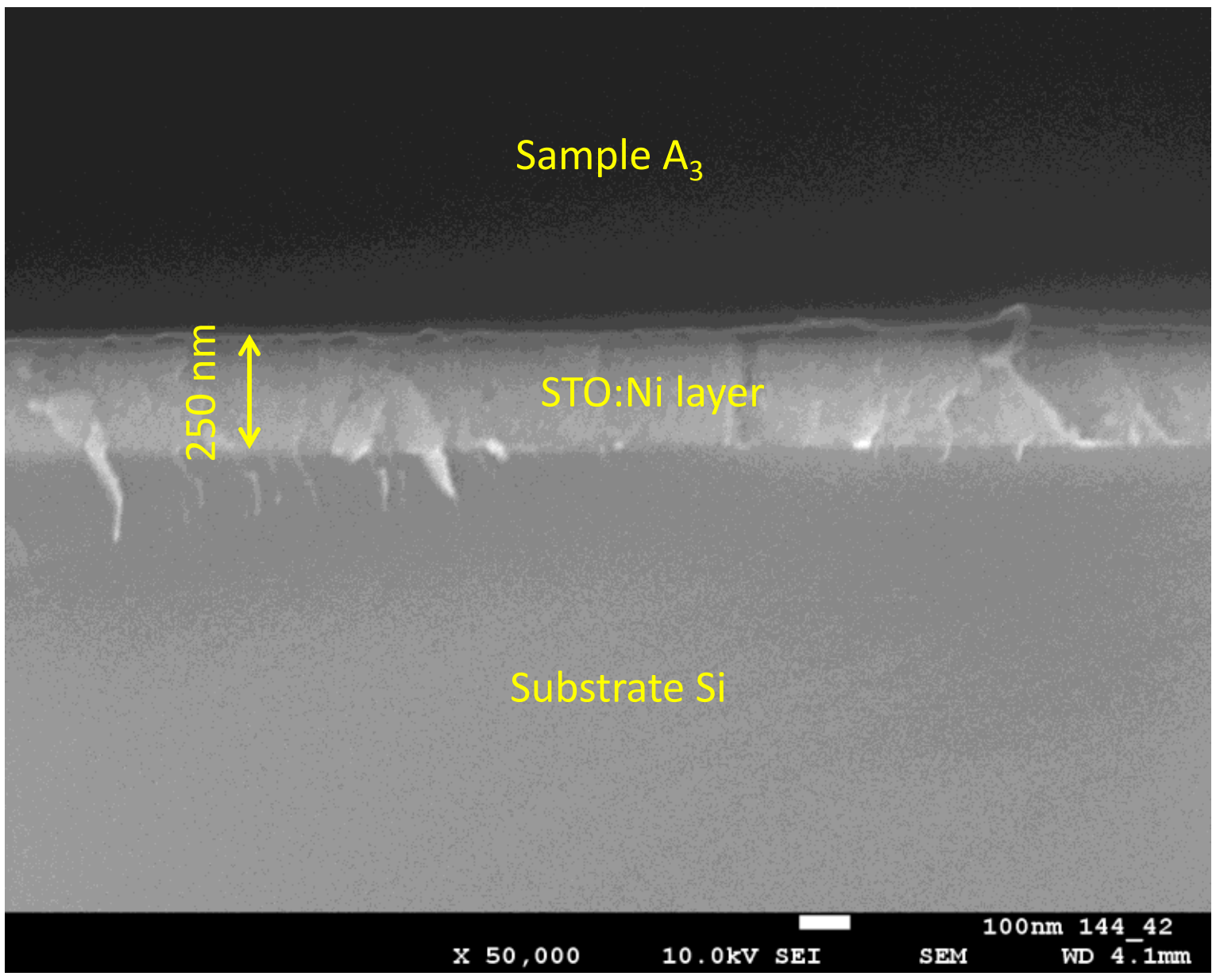

Figure 7 - Perpendicular view of the fracture surface of the $A_3$ sample, magnification 50 000x, layer thickness 200 nm.

Figures 7 and 8 demonstrate the studied layer of the A3 sample. Figure 7 shows a perpendicular view of the fracture surface. From this view, the layer thickness (250nm) can again be determined and the columnar texture of the layer growth can also be seen here. The epitaxy of this sample to the Si substrate is good. In the figure there is fracture damage to the Si substrate after it has been broken. Figure 8 again shows cracks in the surface of the sample. Furthermore, there are similar formations in the surface of the layer. Given the probable identification of the rhombohedral Ni(TiO3) phase, it can be concluded that this is the phase. The amount of this phase is at the very limit of the detection limit of the method used and is therefore not marked on the diffractogram. In the Si substrate surface, cleavage fracture surfaces are visible, which are connected at the points of epitaxial growth of the thin layer to the grains of the layer.

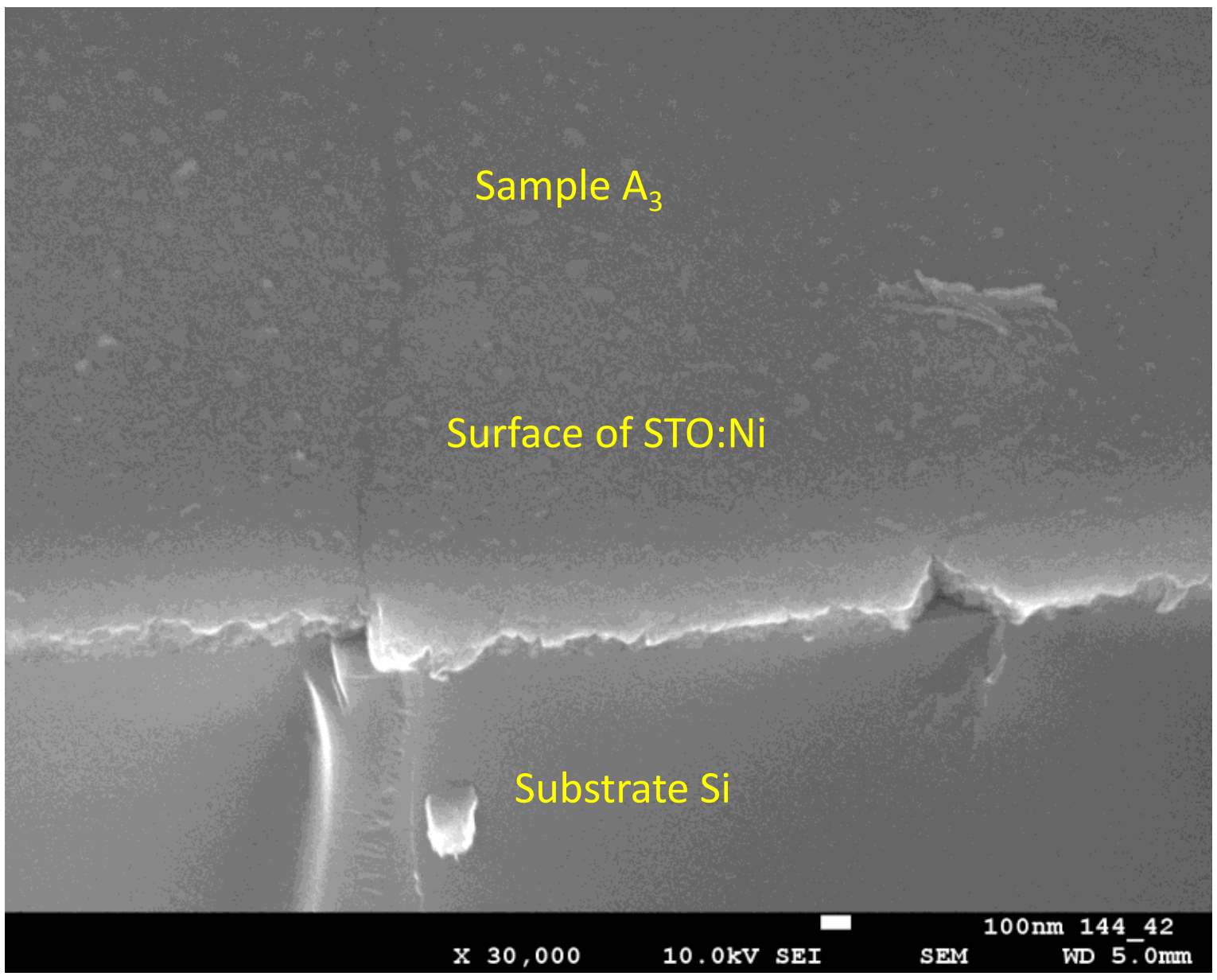

Figure 8 - 45° view, magnification 30 000x, faint cracks in the layer surface visible.

**X-ray photoemission spectroscopy (XPS)**

XPS was used on three selected samples. Samples $A_2$ (2 pellets without $O_2$ flow on deposition), $A_2$ *with* $O_2$ (2 Ni pellets with $O_2$ flow) and $A_3$ (3 Ni pellets without $O_2$ flow) were selected. The experiment was performed on all the elements present in the structure (Sr, Ti, O and Ni). The measured energies of oxygen ions (O 1s) and strontium ions (Sr 3d) had no deviations from the standard energies in the undoped STO structure. Thus, only the differences in the values of titanium (Ti 2) and nickel (Ni 2p) ions, which are crucial for the assessment of the oxygen effect, are described below.

Figure 9 (a) and (b) show the evaluated measured data of the individual elements present in the $A_2$ sample (a) Ti2p and (b) Ni2p). Titanium Ti2p with its doublet is evaluated under a). The binding energy value is 458.0 eV. From [8], it can be found that the binding energy in the STO crystal is 457.4 eV. Thus, it can be seen that the binding energy is affected, albeit minimally. Thus, the binding energy is different and is affected by the addition of Ni dopant. The evaluated record of the binding energy measurement of Ni dopant is given in (b). The value is 855,5 eV. In [9] the value given for Ni2p3/2 is 856.0 eV and according to the XPS database it refers to Ni-O binding. This demonstrates the incorporation of nickel at the titanium position in the oxygen octahedron and the formation of Ni-O bond.

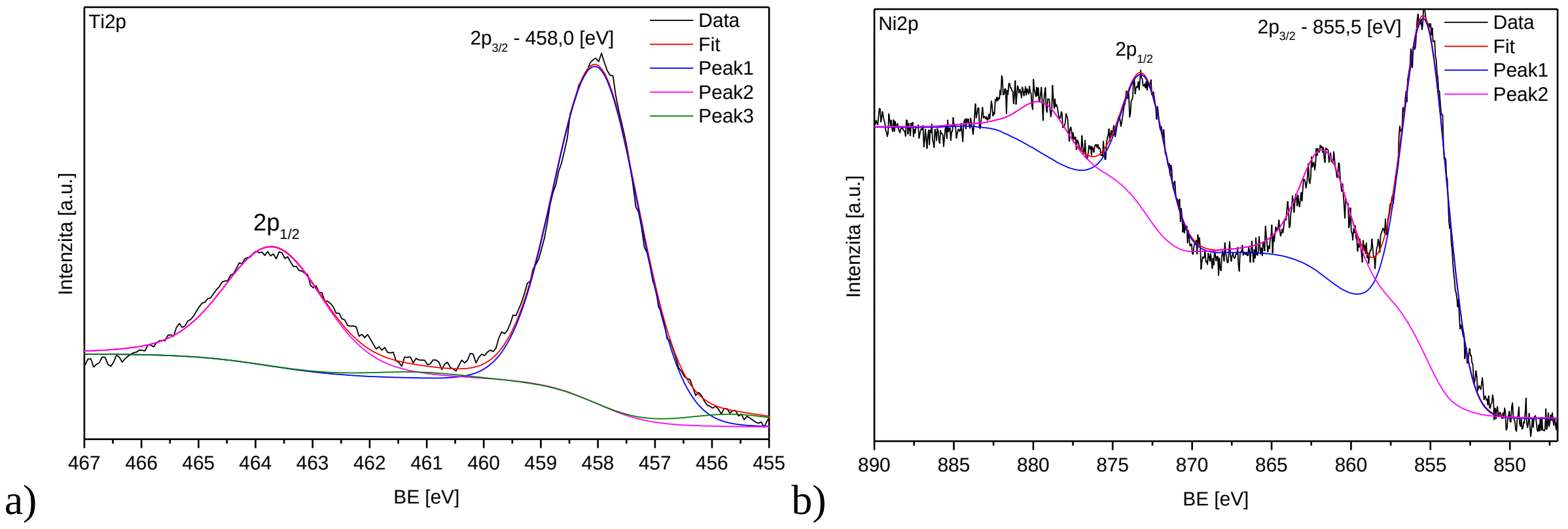

Figure 9 - XPS spectrum of sample $A_2$, beam energy 148.6 eV at room temperature; a) titanium Ti2p and b) nickel Ni2p.

Figure 10 evaluates the measured BE eV binding energies of sample $A_2$ with $O_2$ for the valence layers of a) titanium Ti2p and b) nickel Ni2p.

The value of the binding energy of titanium Ti2p is given under (a). Again, this is a doublet of Ti2p3/2 (corresponding to the Ti3+ ion) and Ti2p1/2 (Ti4+). The value of the binding energy is 458.2 eV. The binding energy in the STO crystal [8] is 457.4 eV. The intensity of the measured binding energy of the nickel dopant Ni2p3/2 is very low. The intensity is likely to be affected by the oxygen flow through the deposition chamber. Due to the very small amount of dopant in the STO lattice and the very weak signal obtained in the measurement, the data were not fit. To improve the signal, measurements with a much longer time would have to be performed, but then the compliance with the same experimental conditions would be affected. The value for pure Ni2p3/2 given in [10] is 852.7 eV. However, from the estimated value from the graph 856.0 eV, the binding is consistent with the assumption of Ni-O binding.

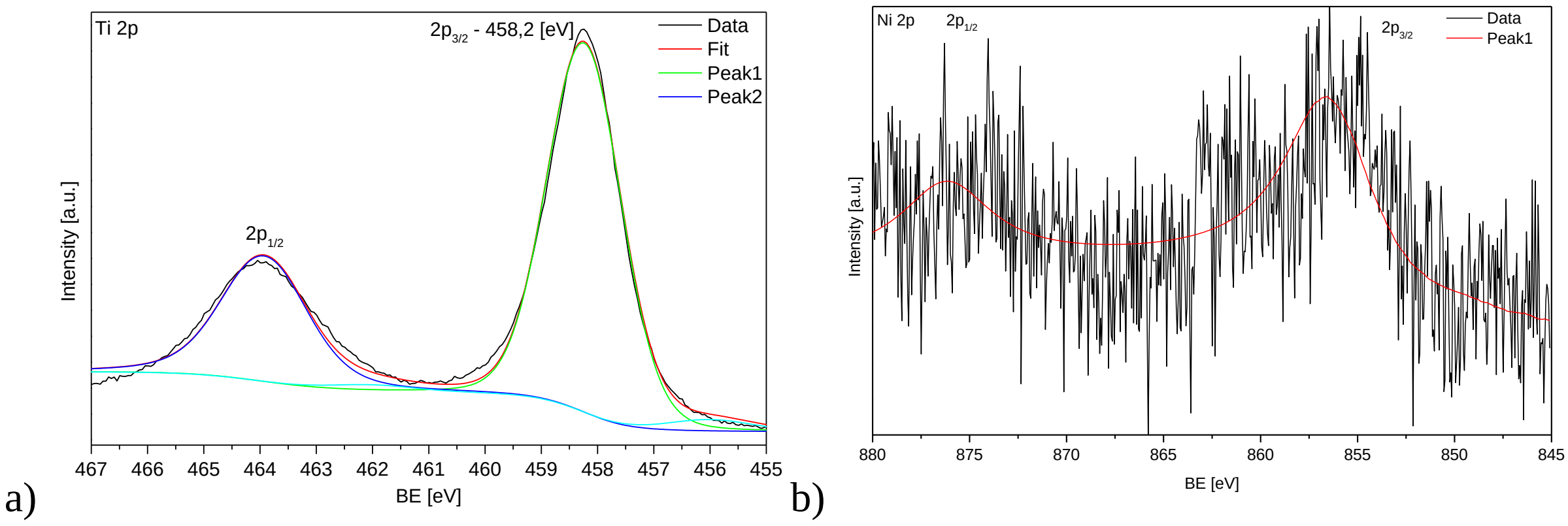

Figure 10 - XPS spectrum of sample $A_2$ *with* $O_2$, beam energy 148.6 eV at room temperature; a) titanium Ti2p and b) nickel Ni2p.

The evaluated measured data of XPS spectra in sample $A_3$ are shown in Figure 11 for a) Ti2p and b) Ni3d. By evaluating the binding energy of Ti, shown in (c), the value of this energy was found to be 457.8 eV. Thus, it can be seen that the influence of the Ti ion in the middle of the oxygen octahedron isac lower. Similarly, evaluation of the binding energy of the Ni dopant suggests that its binding to the oxygen tetrahedron is lower with a value of 855.1 eV, which is closer to the comparative value of 852.7 eV of metallic Ni. Nevertheless, this binding energy is consistent with Ni-O binding [11]. Moreover, the spectrum of Ni2p shows much higher binding energy intensities corresponding to the increased amount of Ni in the STO structure.

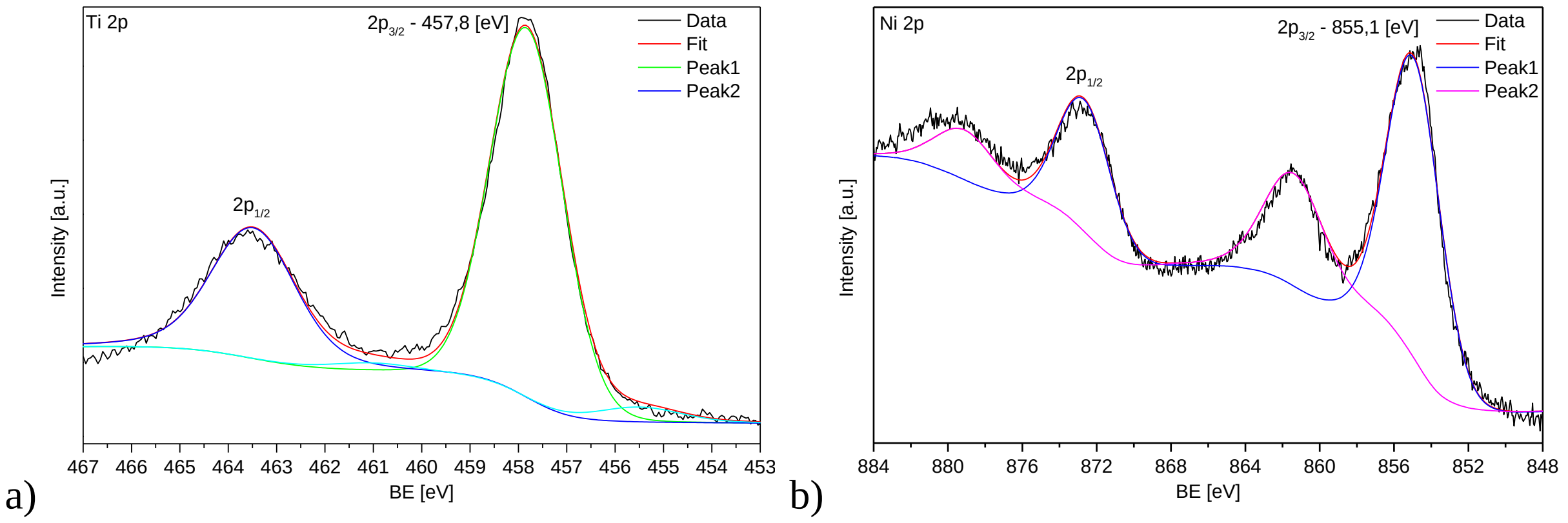

Figure 11- XPS spectrum of sample $A_3$, beam energy 148.6 eV at room temperature; a) titanium Ti2p and b) nickel Ni2p.

## Conclusion

Based on the data obtained from the investigated samples of both series of STO doped with different amounts of nickel with or without $O_2$ flow through the deposition chamber, the following summary can be established.

In the first stage of the experiment, the samples in the as-deposited state were shown to contain only the amorphous phase and the initial crystallization of the STO and TiO phases. It always depended on the amount of dopant used. After the high-temperature experiment, when all samples were annealed at 900°C, crystallization occurred and the amorphous phase disappeared. In the structure of the polycrystalline sample, the STO phase was always the major phase after annealing. When the samples were prepared without the addition of dopant, only the TiO phase was identified as the minority phase. At the same time, the evaluation of these samples showed a very good agreement with the used standard 01-084-0443 from the PDF-2 database, where the difference in the positions of the maximum of the strongest diffraction line of the (110) plane was only 0.08° 2θ. After the addition of dopant and its incorporation into the STO structure, there was a shift of the diffraction lines for all samples compared to the standard. This effect may be due to the addition of Ni. According to the observed XPS data, it can be assumed that dopant atoms replace Ti atoms in the STO ground lattice. This causes a tensile stress and shifts the diffraction line to the left of the standard line. The shift is evident in all diffraction lines of the STO. This is most likely due to the different crystallographic parameters of the two elements. According to the standard 01-084-0443, STO has a cubic lattice with a simple Pm-3m, a = 3.898 [Å]. The crystallographic parameters of Ni dopant are cubic lattice area-centered Fm-3m, a = 3.54 [Å]. The atomic radius of Ti and Ni is 147 [pm] and 124 [pm], respectively.

Figure 4(a) and 4(b) show the effect of the $O_2$ flow rate demonstrated on $A_2$ and $A_2$ with $O_2$. For sample $A_2$ with $O_2$, a greater effect on the increase in crystallinity of the sample was observed. At the same time, Figure 4 b) shows that this sample had less microdeformation inside the lattice. This may be due to less substitution of Ni dopant at the Ti site in the STO lattice.

By evaluating the XPS spectra of samples $A_2$, $A_2$ *with* $O_2$ and $A_3$, it was found that the Ni dopant replaces titanium in the STO lattice, which is in the middle of the oxygen tetrahedron. As a result, ferroelectric phenomena can occur [1], which can be exploited for photocatalytic applications. In the case of comparison of $A_2$ and $A_2$ *with* $O_2$, the difference in the measured data for the doping nickel is noticeable. Both samples have used the same amount of nickel pellets (2 pieces) during deposition, but in the case of sample $A_2$ *with* $O_2$, the air flow through the deposition chamber during deposition has been increased. Thus, the flowing oxygen obviously prevented the incorporation of nickel ions into the STO structure. From the measured data, it can be evaluated that the energies refer indeed to Ni-O bonding, which, however, occurs outside the doped STO:Ni structure. This conclusion is in agreement with the evaluated diffractogram of this sample, where the $Ni(TiO_3)$ phase was detected with the simultaneous formation of the $Sr_5Ni_4O_{11}$ phase. And at the same time, this conclusion is supported by the findings from the SEM examination.

Thus, two conclusions can be drawn from this experiment. First, the oxygen flow during sample deposition improves the subsequent crystallization. However, its presence prevents the incorporation of dopant into the STO lattice by substitution for Ti. This results in the formation of other phases ($Ni(TiO_3)$, $Sr_5Ni_4O_{11}$) whose presence is not favorable for the intended applications.

**Acknowledgements**

This publication was supported by the project Quantum materials for applications in sustainable technologies (QM4ST), funded as project No. CZ.02.01.01/00/22_008/0004572 by Programme Johannes Amos Commenius, call Excellent Research.